\documentclass{article}
\usepackage{graphicx}

\usepackage[utf8]{inputenc} % allow utf-8 input
\usepackage[T1]{fontenc}    % use 8-bit T1 fonts
\usepackage{hyperref}       % hyperlinks
\usepackage{url}            % simple URL typesetting
\usepackage{booktabs}       % professional-quality tables
\usepackage{amsfonts}       % blackboard math symbols
\usepackage{nicefrac}       % compact symbols for 1/2, etc.
\usepackage{microtype}      % microtypography
\usepackage{lipsum}		% Can be removed after putting your text content
\usepackage{graphicx}
\usepackage{doi}
\usepackage{multicol}
\usepackage{graphicx}
\usepackage{subfig}
\usepackage{lipsum}
\usepackage{comment}
\usepackage{biblatex}
\usepackage{algorithm}
\usepackage{algorithmicx}
\usepackage[noend]{algpseudocode}
\usepackage{amsmath} % for math symbols and environments

\title{Exploring Softly Masked Language Modelling for Controllable Symbolic Music Generation}
\author{Nicolas Jonason \\ njona@kth.se \and Bob L.T. Sturm \\ bobs@kth.se}
\date{Version 1.1}

\newcommand\rollwidth{.48}

\newcommand{\pianorollfig}[2]{
\begin{figure}[H]
\centering

\subfloat{%
\includegraphics[width=\rollwidth\linewidth]{#1_0.png}%
}\hfill
\subfloat{%
\includegraphics[width=\rollwidth\linewidth]{#1_1.png}%
}\\
\subfloat{%
\includegraphics[width=\rollwidth\linewidth]{#1_2.png}%
}\hfill
\subfloat{%
\includegraphics[width=\rollwidth\linewidth]{#1_3.png}%
}
\caption{#2}
\end{figure}
}

\begin{document}

\maketitle

\begin{abstract}

This document presents some early explorations of applying Softly Masked Language Modelling (SMLM) to symbolic music generation.
SMLM can be seen as a generalisation of masked language modelling (MLM), where instead of each element of the input set being either known or unknown, each element can be known, unknown or partly known. 
We demonstrate some results of applying SMLM to constrained symbolic music generation using a transformer encoder architecture.
Several audio examples are available at {\href{https://erl-j.github.io/smlm-web-supplement/}{https://erl-j.github.io/smlm-web-supplement/}}
\end{abstract}

\section{Background}

Symbolic music generation systems aim to assist in the composition of music. One challenge is to make systems that are highly controllable so that humans can interactively develop musical ideas.\cite{,pachet_markov_2011,hadjeres_deepbach_2017,huang_counterpoint_2019,hadjeres_piano_2021,chang_variable-length_2021,ens_mmm_2020, hadjeres_anticipation-rnn_2020, zeng_musicbert_2021}
One way to exert control is to assign values to note attributes such as pitch, onset time or duration.
Rather than directly assigning values to note attributes, SMLM allows us constrain the note attributes to a set of values that the model is then able to choose from
Importantly, SMLM takes into account constraints across all notes in the composition during generation of each note. This means that the model is able to account for the constraints imposed on other notes before generating a particular note.
We believe that constraining note attributes to a set of values rather directly assigning values improves the controllability of symbolic music generation systems.
For example, given a simple pitch/onset/attribute-representation we can make the composition use a particular scale by constraining pitch, make it follow a particular rhythm by constraining the onsets or perform imputation of areas of the piano roll.
\begin{figure}[H]
    \centering
    \includegraphics[scale=0.22]{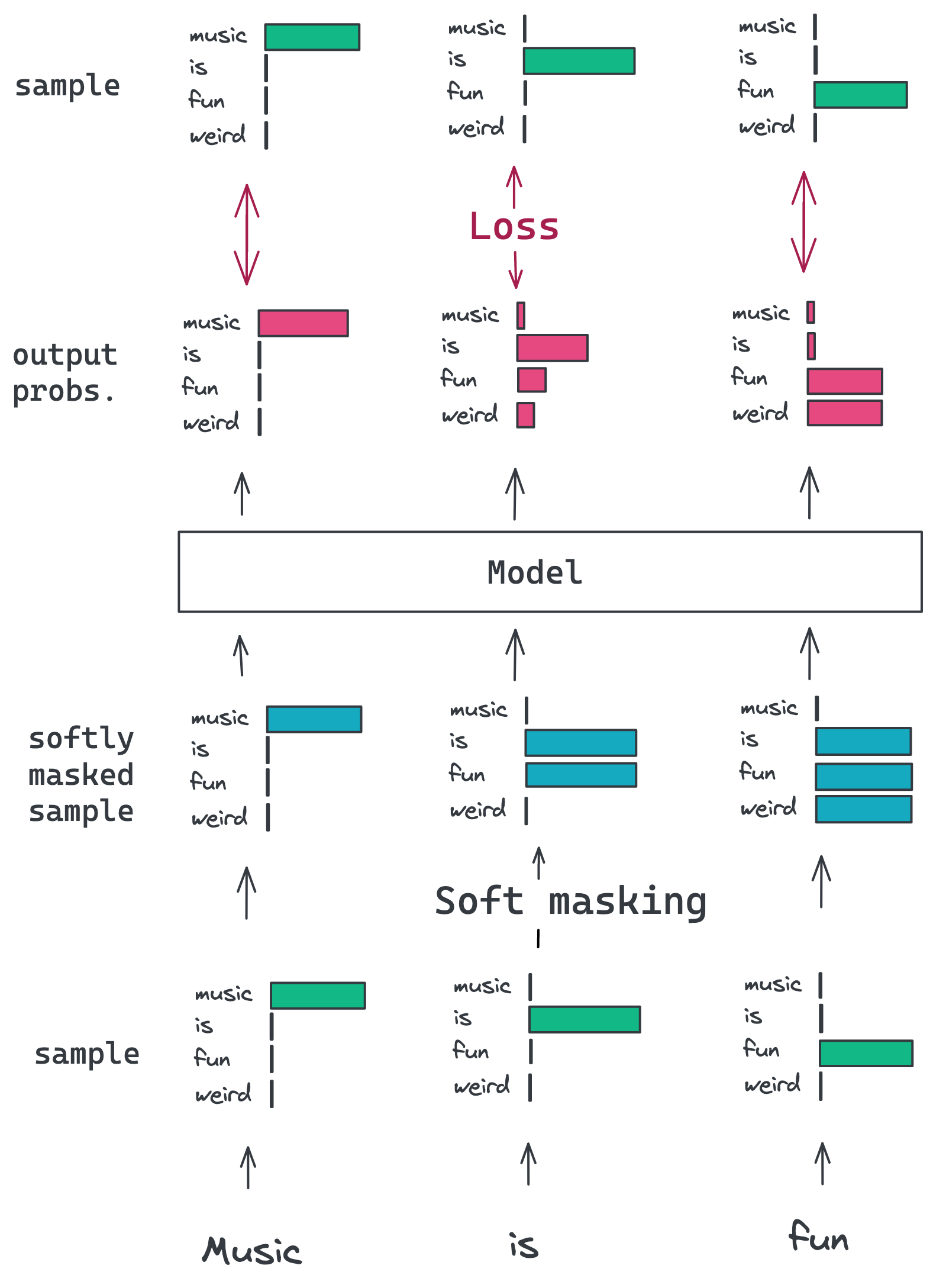}
    \caption{Illustrating example of SMLM training on natural language. At the bottom, an input is "softly masked" by adding confounding information about the identity of each token. The prediction model processes the softly masked input and outputs distributions for each token. Finally, the loss is computed by comparing the output probabilities to the ground truth.
    }
    \label{fig:smlm}
\end{figure}

\section{Softly Masked Language Modelling}

%In this section, we will explain SMLM, then give an example of an SMLM applied to text generation.

Given a size-$T$ input set $X = \{x_1, x_2, \dots, x_T\}$ with element-wise prior information, $p_i$ for each element $x_i$, the softly masked language modeling (SMLM) task aims to minimize the negative log-likelihood of the elements given the prior information:

%In a categorical smlm

\begin{equation}
\mathcal{L}_{\text{SMLM}}(\theta) = - \sum_{i=1}^{T} \log P(x_i | p_1, p_2, \dots, p_T; \theta),
\end{equation}

In order to train a SMLM, we need to define a process which takes an input set $X = \{x_1, x_2, \dots, x_T\}$ and extracts the element-wise prior information 
$P = \{p_1, p_2, \dots, p_T\}$ We refer to this process as \emph{softly masking} the input $X$.

Let's illustrate this with an example in natural language generation. Our input will consist of one hot vectors $X = \{x_1, x_2, \dots, x_T \}$ representing each token in the input. We choose the following soft masking process : starting from a one hot vector $x_i$, we add new, ```false``` activations so that it becomes a multi-hot vector. This is illustrated in the bottom of figure \ref{fig:smlm}.

%In the co¨ntext of a set of elements composed of categorical attributes, we can produce the element-wise prior information by adding new, "false" activation's to the input vector representing each element attribute (see example in figure \ref{fig:smlm}). We call this process \emph{softly masking}.

\section{Training a Softly Masked Language Model On Music Loops}
% Mention masking
% Add details

\subsection{Dataset}
We use a dataset of short excerpts extracted from the MetaMIDI dataset.\cite{ens_metamidi_2021}
We check that the MIDI file contains at most one time signature message and that the time signature is 4/4.
Then, we quantize the note onset times to 16th notes.
We discard all events from MIDI channel 10 (which corresponds to drums).
Finally, we segment the tracks into 64 step excerpts without overlap starting from the first note event of the MIDI file.
During training, we randomly crop the segments so that it only spans 36 pitches.  

\pianorollfig{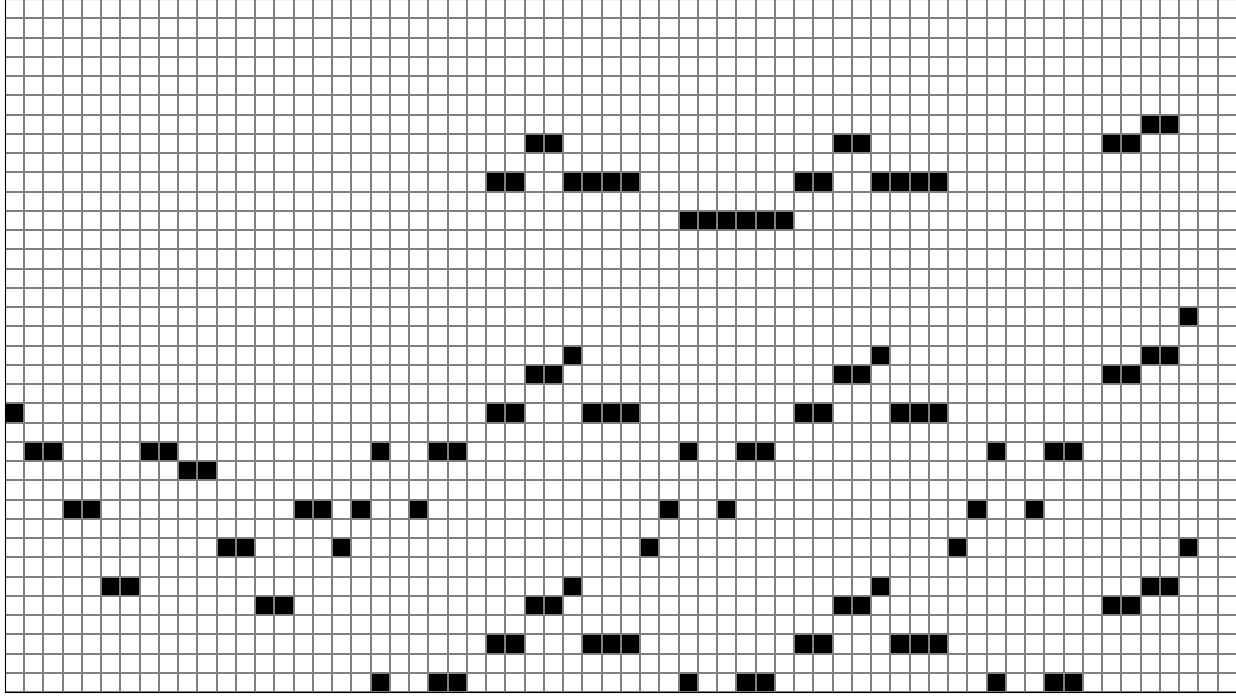}{Random samples from the dataset}

\subsection{Representation}
For simplicity, we consider music as a set of note events with three attributes.
\begin{itemize}
    \item pitch $\in$ {0-35, "undefined"}
    \item onset time $\in$ {0-63, "undefined"}
    \item duration $\in$ {1-63, "undefined"}
\end{itemize}
This representation is similar to the OctupleMIDI representation \cite{zeng_musicbert_2021}.

One implication of our representation is that there exists invalid combinations of element attributes. This is the case when one attribute is known to be undefined while the others attribute are known to not be undefined. To address this issue, we use a normalisation function during every forward pass of the model.

\subsection{Masking scheme}

During training, we generate masks in two stages. In the first stage, we mask
particular parts of the vocabulary across all masked elements uniformly. In the second stage, we randomly mask parts of of the vocabulary across each masked element independently. We randomly sample the number of elements to be masked in each stage. 

\subsection{Architecture}
We encode each element attribute with attribute specific encoders.
These encoders are implemented as fully connected layers with bias.
The element attributes are pooled by summing the embedding of each attribute to form an element embedding. 
For our main block, we use a neural network with the transformer encoder architecture \cite{vaswani_attention_2017,devlin_bert_2019}.
We use a hidden size of 768, 8 layers, and 8 attention heads.
We decode the main block output using attribute specific decoders.
These decoders are implemented as fully connected layers with bias. The decoders output logits for each attribute.
As a final step we process the logits so that they are guaranteed to be consistent with the constraint. We achieve this by subtracting a large number from the logits of the events that are not permitted by the constraint.

\subsection{Training}

We use a learning rate of 1e-4 with a learning rate decay of 0.95. We use a batch size of 384. 
Training lasted 48 hours or 77 epochs using a NVIDIA GeForce RTX 3090. Training loss and validation losses were approximately equal ($\sim$1.80 / 1.75) at stopping time.

\subsection{Generation}

In order to generate note that satisfy a constraint, we express our constraint in the form of a mask. For example, if we want to only allow notes of duration 4 and 8 or no note at all, we fully mask pitch and onset attributes and only mask the value 4, 8 and undefined for the duration attribute. We then generate the composition by sampling attributes from notes in random order.
%We can speed up generation by first determining the number of inactive notes after the first forward pass. We determine if a particular note will be inactive by sampling the ``undefined" value of any of its three attributes.

\section{Generation examples}

All samples are generated with temperature of 1.0 and top-p with $p=0.9$ unless indicated otherwise. Generated areas are shown in green. Audio is available at {\href{https://erl-j.github.io/smlm-web-supplement/}{https://erl-j.github.io/smlm-web-supplement/}}.

\subsection{Unconditional generation}

\pianorollfig{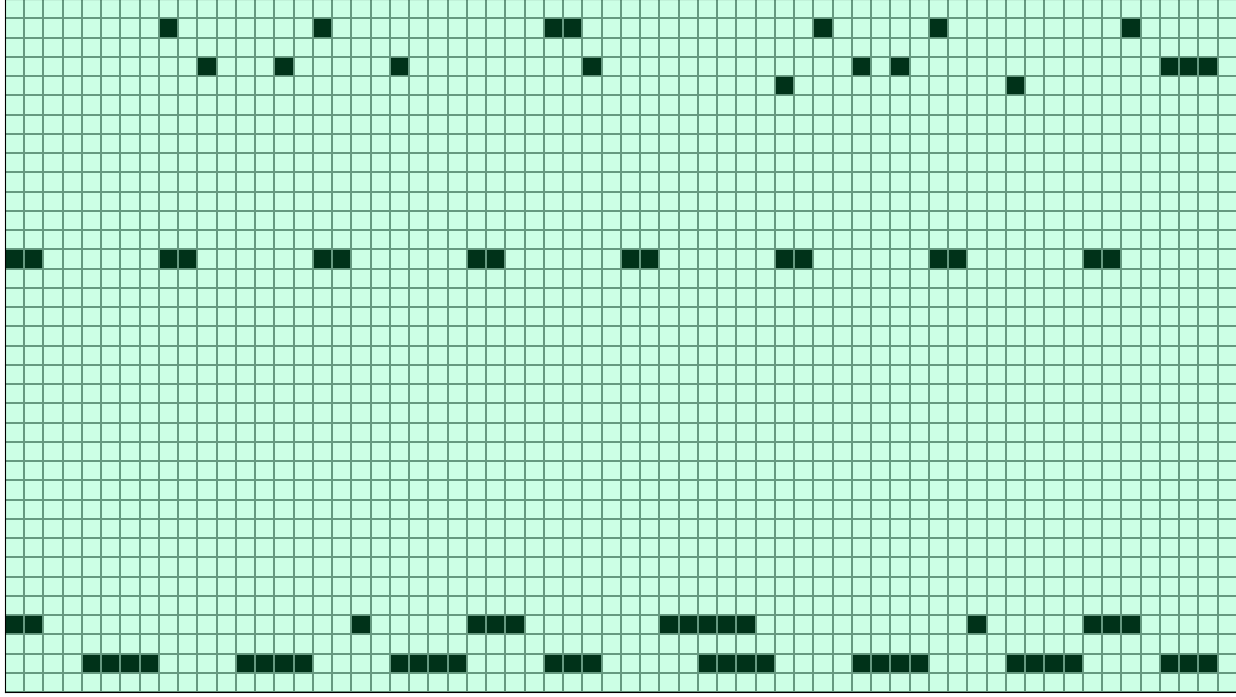}{Generation without any constraints}

\subsection{Imputation}

\pianorollfig{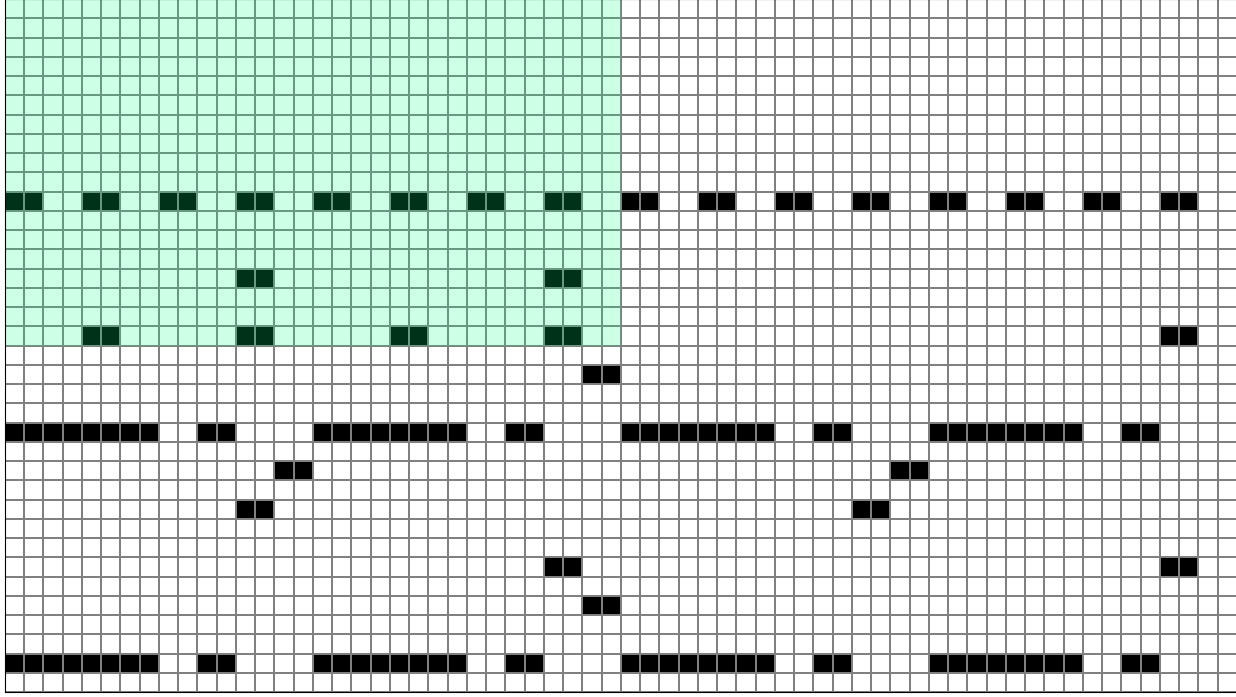}{
Imputation of a rectangular area in pitch-time space.
Generated with temperature=0.75 and p=0.9.
}

\pianorollfig{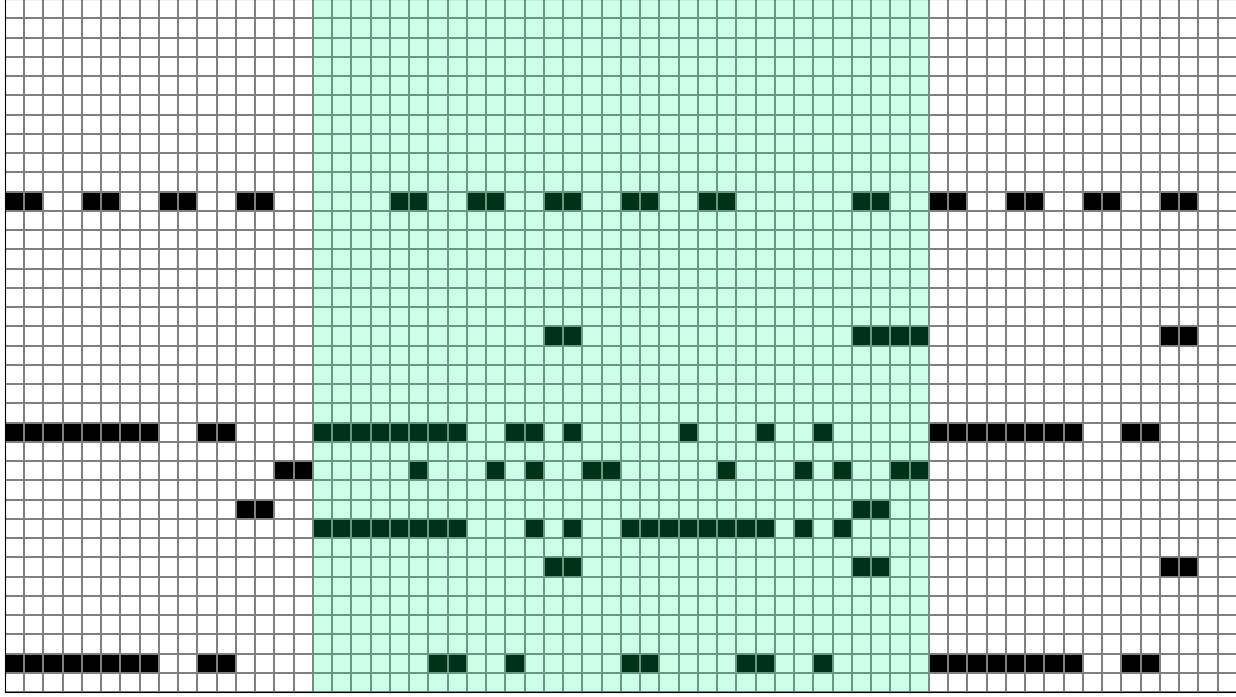}{
Imputation of a rectangular area in pitch-time space.
Generated with temperature=0.75 and p=0.9.
}

\pianorollfig{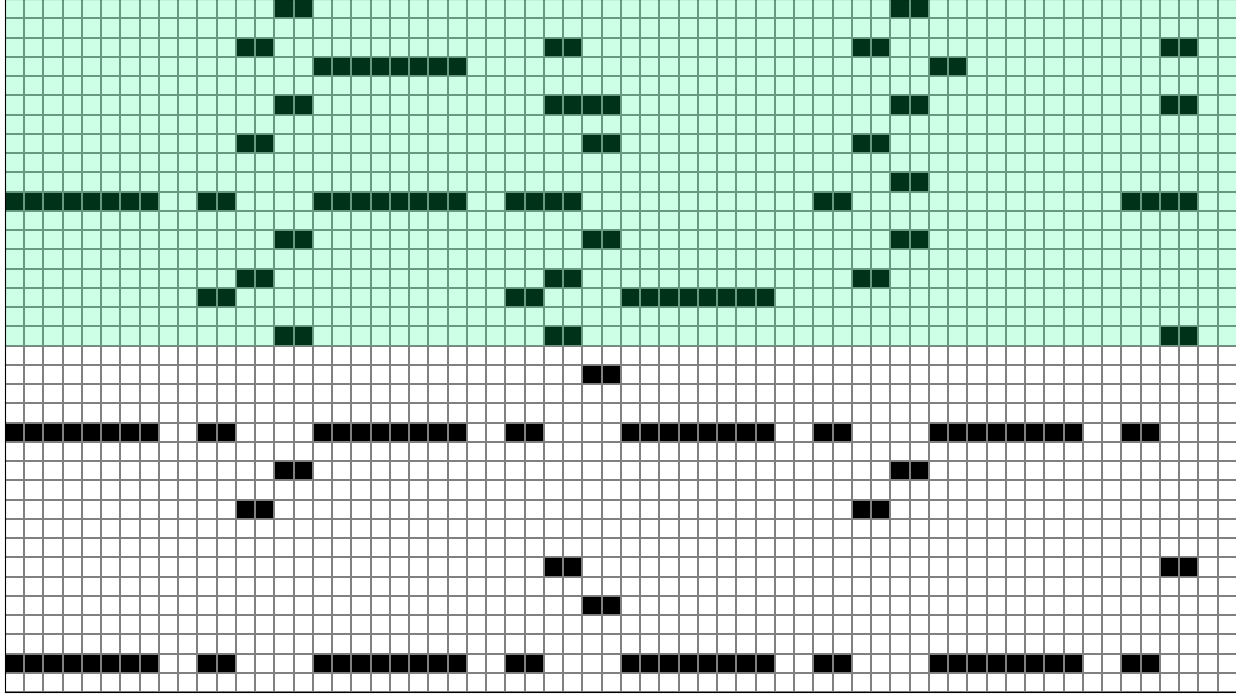}{
Generation of high pitches conditioned on low pitches
}

\pianorollfig{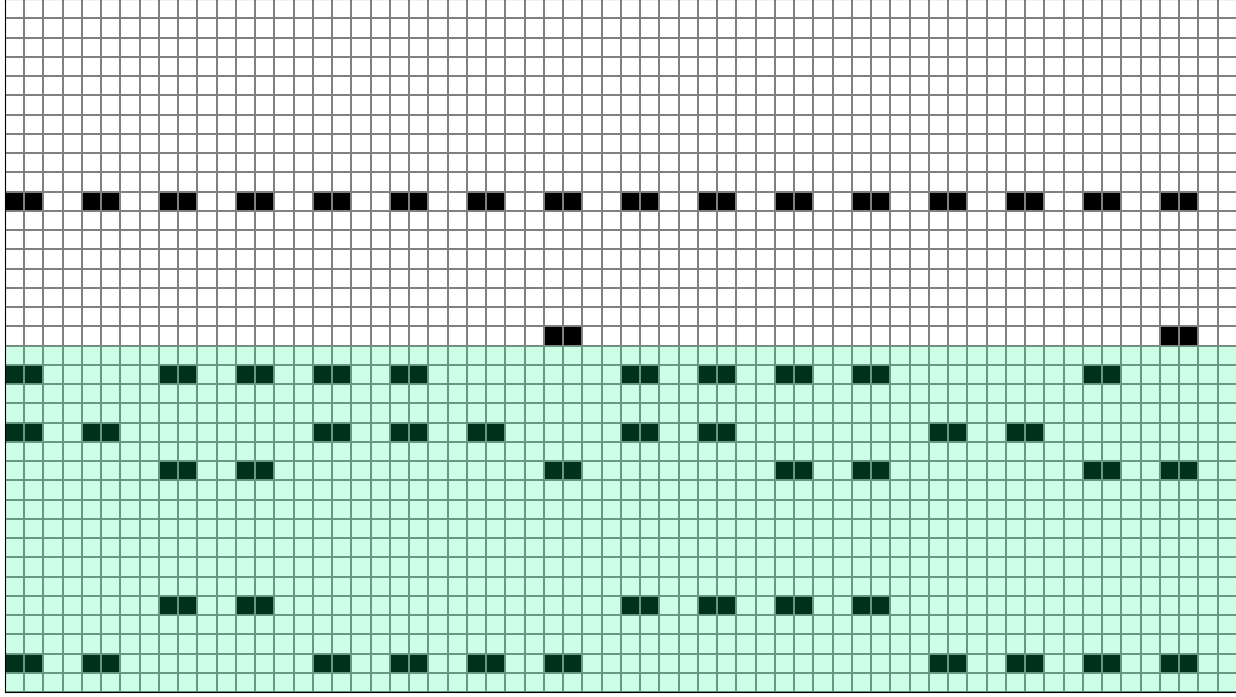}{
Generation of low pitches conditioned on high pitches.
}

\subsection{Pitch control}

\pianorollfig{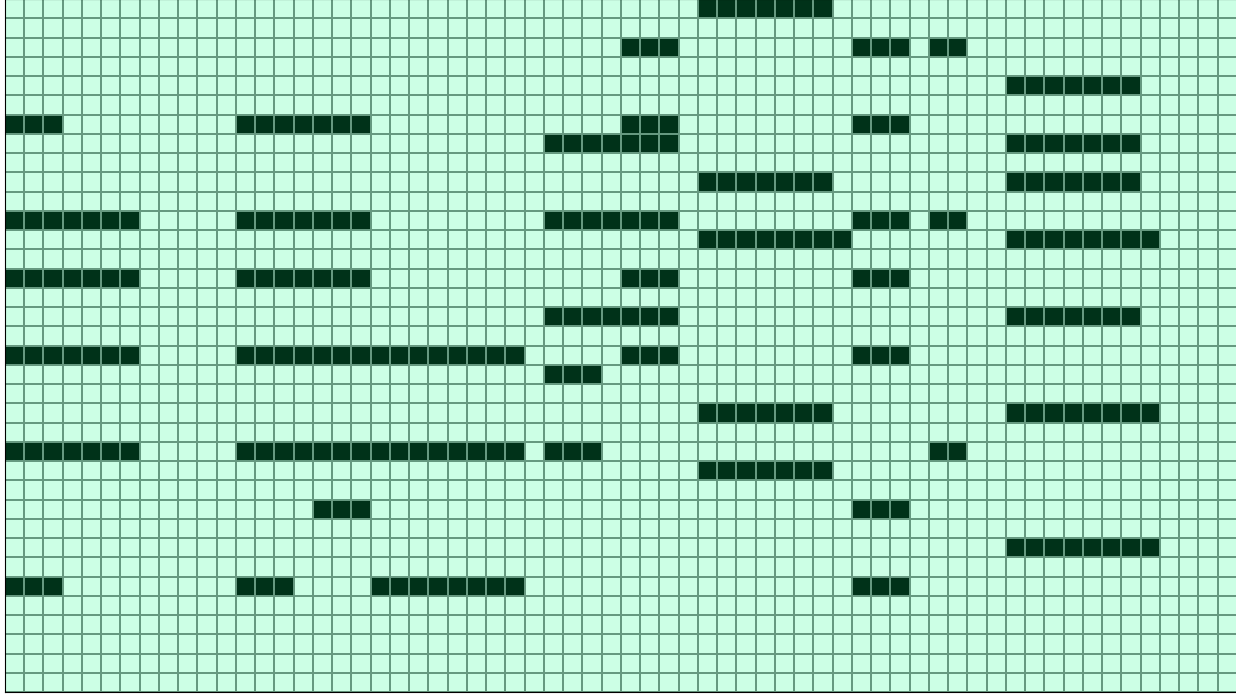}{Generation with pitch constrained to major scale with root at pitch 0}

\pianorollfig{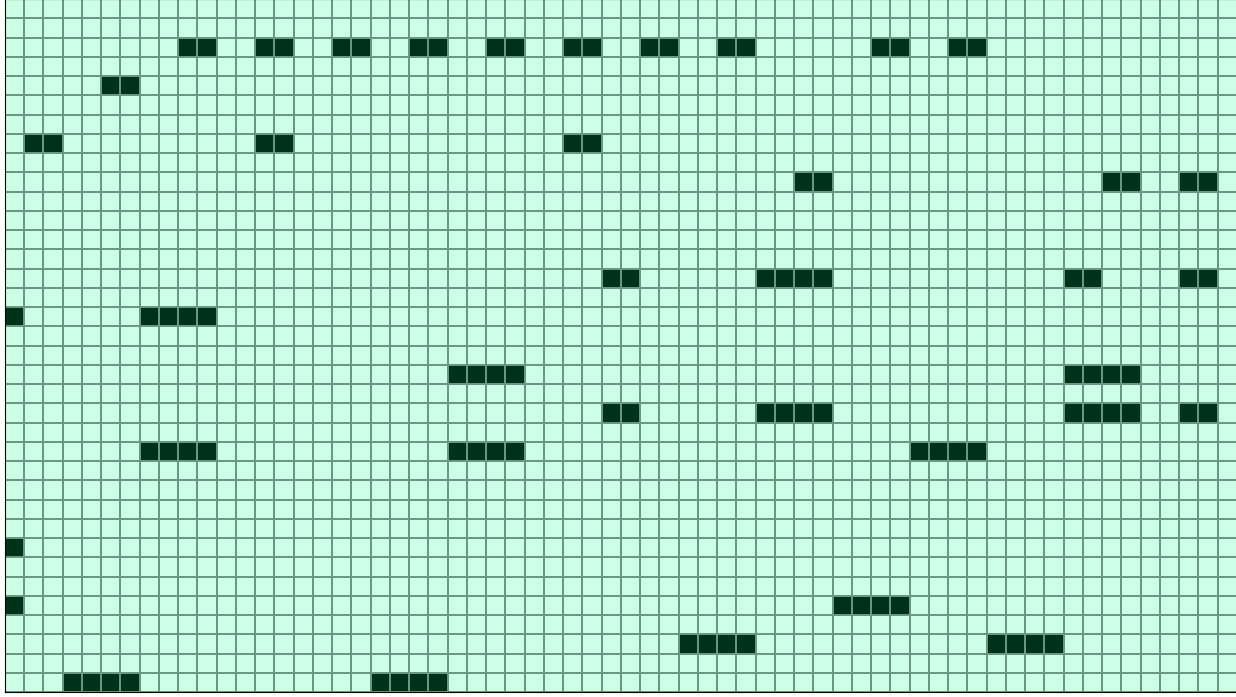}{Generation with pitch constrained to major pentatonic scale with root at pitch 0}

\subsection{Duration control}

\pianorollfig{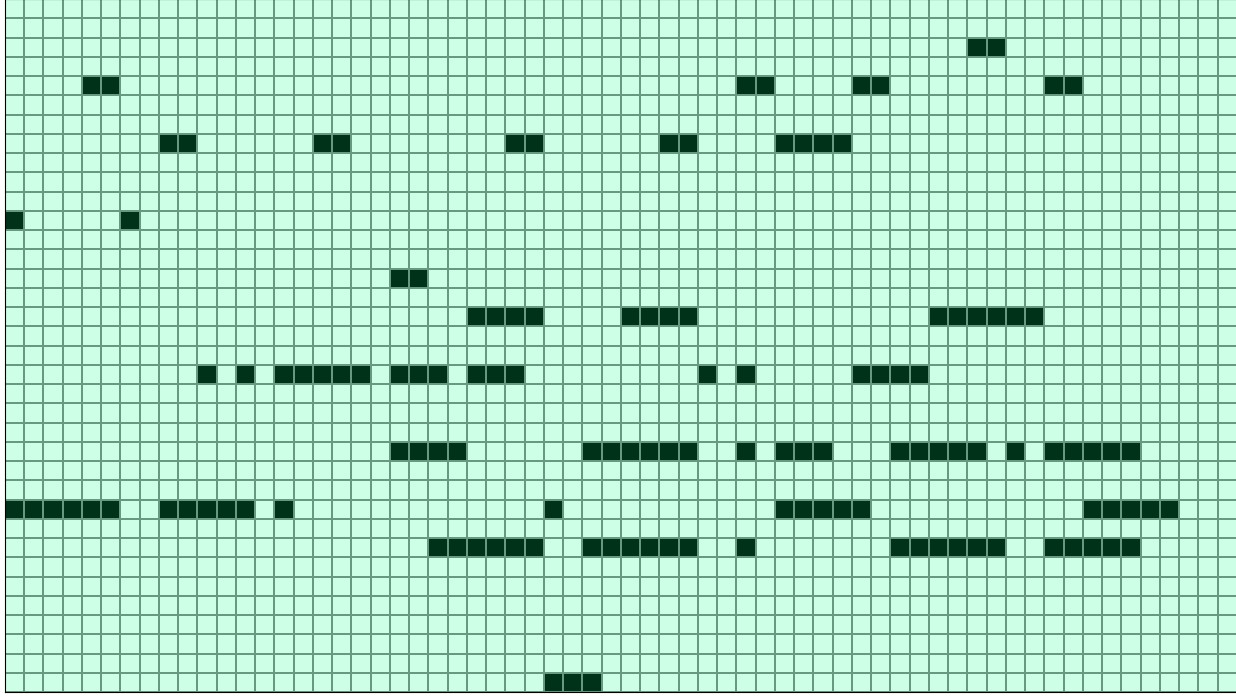}{
Generation with note duration constrained to be less than 8 steps.
}

\pianorollfig{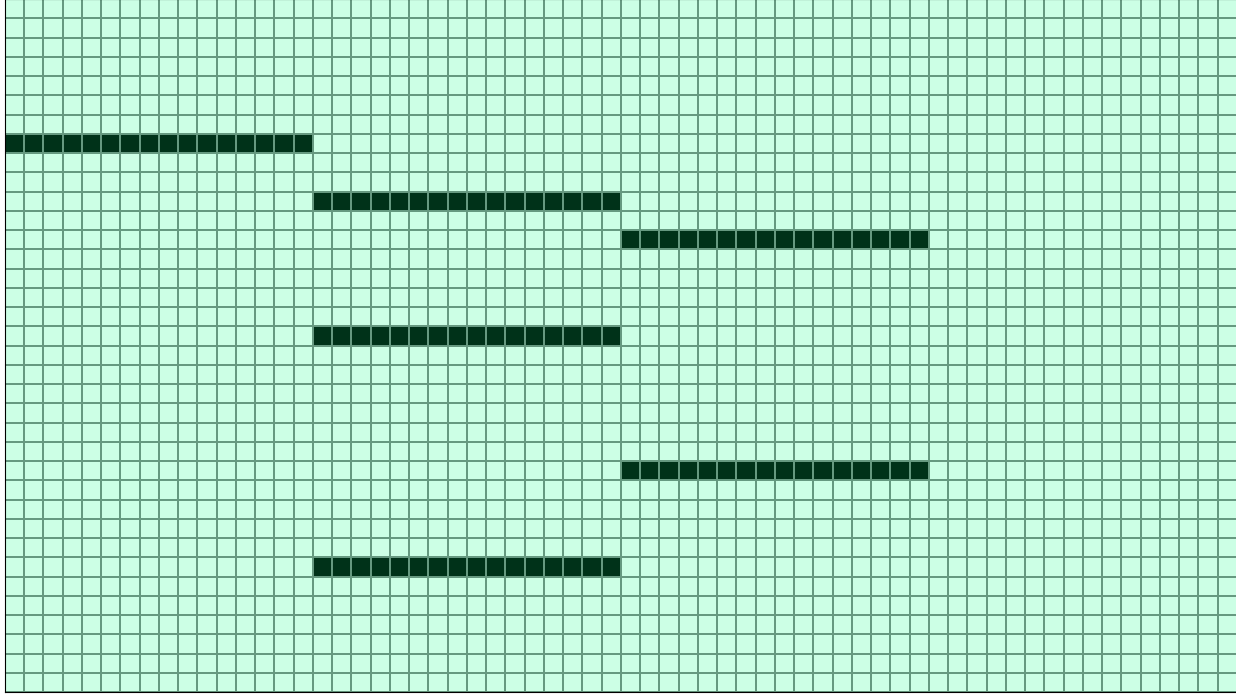}{
Generation with note duration constrained to be more than 8 steps.
}

\subsection{Rhythm control}

\pianorollfig{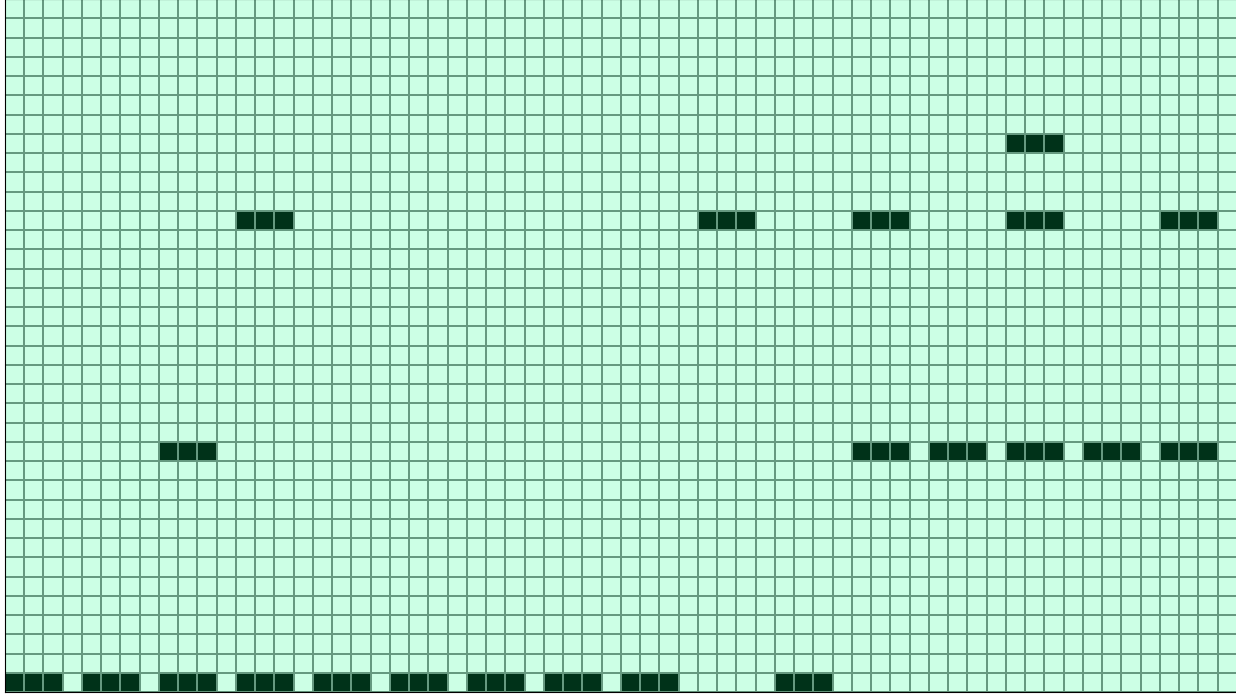}{
Generation with onset time constrained to every 4 steps
}

\pianorollfig{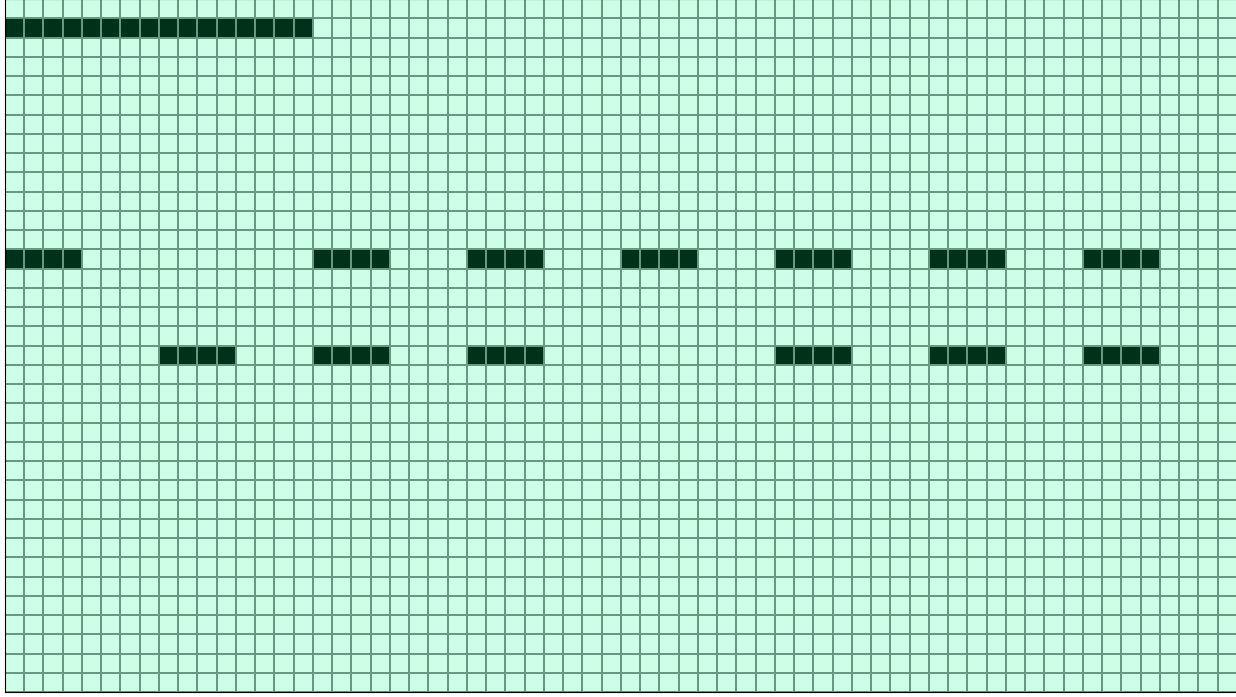}{
Generation with onset time constrained to every 8 steps
}

\pianorollfig{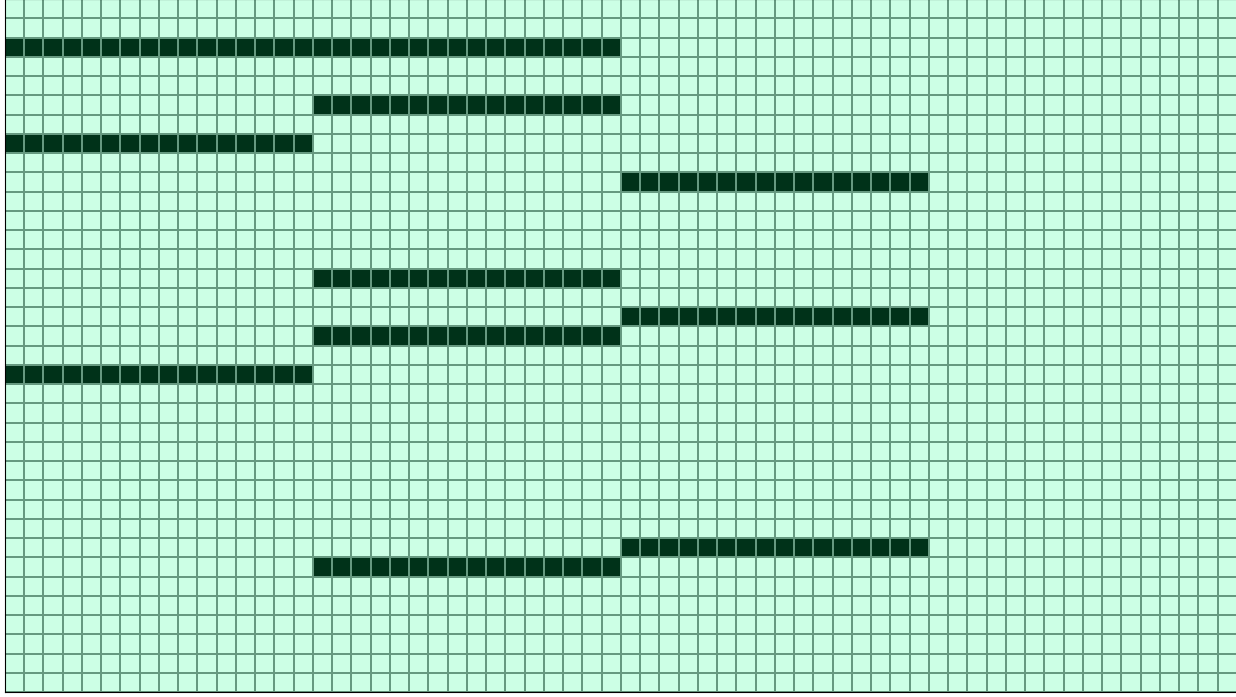}{
Generation with onset time constrained to every 16 steps
}

\subsection{Combining constraints}

\pianorollfig{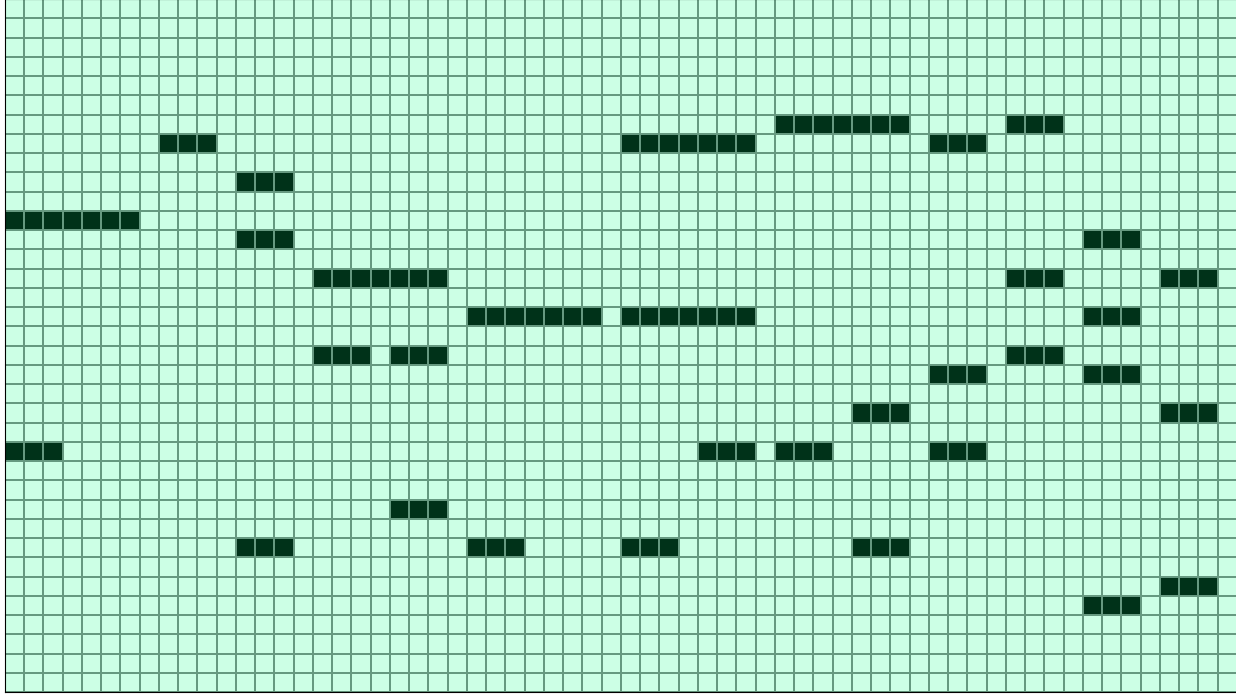}{Generation with multiple constraints. Pitch is constrained to major scale with root pitch 0, duration is set to be between 2 and 7, onsets are constrained to every 4 steps. Minimum number of notes is set to 16, maximum number of notes is set to 64. }

\section{Future work}

In future work we aim to: 1) Expand the representation in order to apply SMLM to a wider range of musical tasks; 2) Explore ways to improve generation quality and generation speed; 3) Perform qualitative and quantitative evaluation of the system.

\section{Acknowledgments}

We thank Gustav Eje Henter and Luca Casini for valuable discussions during this work.

This work is an outcome of MUSAiC, a project that has received funding from the European Research Council under the European Union’s Horizon 2020 research and innovation program (Grant agreement No. 864189).

\printbibliography
\end{document}